\documentstyle[aps,prl,floats,epsf]{revtex}
\begin{document}
\draft
\renewcommand{\theparagraph}{\Alph{paragraph}}
\twocolumn[\hsize\textwidth\columnwidth\hsize\csname@twocolumnfalse%
\endcsname

\title{First order phase transition in a 1+1-dimensional
       nonequilibrium wetting process\\}

\author{H. Hinrichsen$^{(1,2)}$, R. Livi$^{(3)}$, 
        D. Mukamel$^{(1)}$, and A. Politi$^{(4)}$\\[-2mm]$ $}

\address{$^{1}$
        Department of Physics of Complex Systems,
        Weizmann Institute of Science, Rehovot 76100, Israel\\
        $^{2}$
        Max-Planck-Institut f\"ur Physik komplexer Systeme,
        N\"othnitzer Stra\ss e 38, 01187 Dresden, Germany\\
        $^{3}$
        Dipartimento di Fisica, Universit\'a, INFM and INFN, 
	50125 Firenze, Italy\\
        $^{4}$
        Istituto Nazionale di Ottica and INFM-Firenze, 
	50125 Firenze, Italy}

\date{August 09, 1999}
\maketitle
\begin{abstract}
A model for nonequilibrium wetting in 1+1 dimensions is
introduced. It comprises adsorption and desorption processes
with a dynamics which generically does not obey detailed balance.
Depending on the rates of the dynamical processes the wetting
transition is either of first or second order. It is found that
the wet (unbound) and the non-wet (pinned) states coexist and
are {\it both thermodynamically stable} in a domain of the dynamical
parameters which define the model. This is in contrast with
equilibrium transitions where coexistence of thermodynamically
stable states takes place only on the transition line.
\end{abstract} \vspace{2mm}

]
%
%

Wetting phenomena are observed in a large variety
of physical systems in which a planar substrate is
exposed to a gas phase. The interactions between 
the substrate and the molecules of the gas phase 
lead to the formation of a liquid film close to the 
surface~\cite{WettingReview}. 
By changing the temperature or
the partial pressure of the gas, such systems may
exhibit a wetting transition from a phase where
the thickness of the film stays finite to a
phase where the film grows and eventually reaches
a macroscopic size. 

Much is known about equilibrium wetting 
transitions both theoretically and experimentally.
Theoretically, they have been modeled by considering
the binding process of a fluctuating interface to a substrate.
By using transfer matrix formulation, it has been found
that in 1+1 dimensions, when the interaction between 
the interface and the substrate is short range, 
the depinning transition is continuous~\cite{TransferMatrix}. 
However, when algebraically decaying long range interactions 
are taken, the transition could become first order {\cite{ZLK88}}.
On the other hand, in 2+1 dimensions the wetting
transition is generally expected to be first order.
However, disorder effects,
such as those existing in porous media, could make
the transition continuous~\cite{BKD98}.

When passing to non-equilibrium phenomena, 
most of the efforts have been 
made in the characterization of the 
wet phase, namely the description
of the growth process far away from the
substrate. Various models 
have been introduced to account
for observed scaling behaviour in 
homogeneous as well as disordered
environments~\cite{BarabasiStanley95}. 

Concerning the transition itself, 
a simple 1+1 dimensional model has recently
been introduced where a continuous transition 
is observed in the absence of
any binding force between the interface and the 
substrate~\cite{PreviousPaper}.
It would be of interest to study the nature of the
transition as the interaction between the interface
and the substrate is varied.
 
Motivated by the knowledge of the 
behavior of equilibrium systems, we
generalize the model introduced 
in Ref.~\cite{PreviousPaper} by adding an attractive
binding force between substrate
and surface layer.  
We find that even {\it short range} attractive
interaction is capable of making the transition
first order.  
In fact, for a suitable 
choice of the parameter values 
the model satisfies detailed balance 
and thus we can test our numerics
against theoretical predictions in the 
context of equilibrium wetting. 
Additionally we can explore the wider
parameter space to clarify the 
role of the binding force under
nonequilibrium conditions.

The model of Ref.~\cite{PreviousPaper} is 
characterized by an adsorption rate $q$
and a desorption rate $p$ and exhibits a {\em continuous}
wetting transition at a certain threshold $q_c(p)$. 
The transition is related to an unpinning process of
an interface from a substrate, which may be described 
in general by a Kardar-Parisi-Zhang (KPZ) equation in 
a hard-core potential~\cite{KPZWall}. The additional 
short-range force is introduced by modifying the
growth rate $q_0$ at zero height. 
Thus, for $q_0 <q$ ($q_0 > q$), 
there is an attractive (repulsive) 
interaction between the substrate and 
the bottom layer. We find that a sufficiently 
strong attractive interaction modifies 
the nature of the unbinding 
transition, making it first order.
Moreover, we observe that pinned 
and moving states {\it coexist 
as thermodynamically stable states} in a 
whole region of the parameter 
space rather than on a line,
as in the case of equilibrium 
transitions. This kind of behavior has 
been observed in the past in other 
non-equilibrium models~\cite{Toom80,Bridge}
and should be a generic feature of wetting transitions
under nonequilibrium conditions.

\vspace{2mm}
\paragraph{Definition of the model: }
The model is defined in terms of growth of a one-dimensional 
interface on a lattice of $N$~sites with associated 
height variables $h_i=0,1,\ldots,\infty$ and periodic 
boundary conditions. We consider a {\em restricted} 
solid-on-solid (RSOS) growth process, where the 
height differences between neighboring sites can
take only values $0,\pm1$. In addition, a 
hard-core wall at zero height is introduced. 
The model depends on three parameters 
$q,q_0$, and $p$. It evolves by random sequential 
updates, i.e., in each update attempt 
a site~$i$ is randomly selected and one of the 
following processes is carried out:\\[-1mm]

\noindent --  adsorption of an adatom with 
	probability $q_0\;\Delta t$ at the 
	\\ \indent 
	bottom layer $h_i=0$ and probability $q\;\Delta t$
	at higher 
	\\ \indent layers $h_i>0$:
        \begin{equation} \label{Process1}
        h_i \rightarrow h_i+1 \, ,  \end{equation}
        
\noindent -- desorption of an adatom from the edge of a terrace \\
        \indent with probability $1\;\Delta t$:
        \begin{equation} \label{Process2} 
        h_i \rightarrow \min(h_{i-1},h_{i},h_{i+1}) \, ,
        \end{equation}
        
\noindent -- desorption of an adatom from the interior of a terrace \\
        \indent with probability $p\;\Delta t$:
        \begin{equation}  \label{Process3}
        h_i \rightarrow h_i-1 \quad \mbox{if} \quad
        h_{i-1}=h_{i}=h_{i+1} > 0 \, . \end{equation}
A process is carried out only if the resulting interface 
height $h_i$ is non-negative and does not violate the RSOS
constraint $|h_i-h_{i\pm 1}| \leq 1$. The time increment
per sweep (N attempted updates) is $\Delta t \leq 1/\max(1,q_0,q+p)$.

The phase diagram for the case $q_0=q$ has been studied 
in~\cite{PreviousPaper,AEHM}, where a continuous wetting transition
was found. Clearly, the moving state is not affected by $q_0$ and
thus the transition line above which it is stable remains
unchanged. However, the stability of the pinned state
strongly depends on $q_0$, modifying the phase diagram
and the nature of the wetting transition.
In order to gain some insight into the mechanism leading to
first-order transition, we first consider the $p=1$ case.
Here detailed balance is obeyed~\cite{PreviousPaper},
wherefore the transition can be described in the 
framework of equilibrium statistical
mechanics~\cite{Equilibrium}.
We then consider the case $p\neq 1$ numerically.

\vspace{2mm}
\paragraph{The case $p=1$: }
%
%
\begin{figure}
\epsfxsize=85mm
\centerline{\epsffile{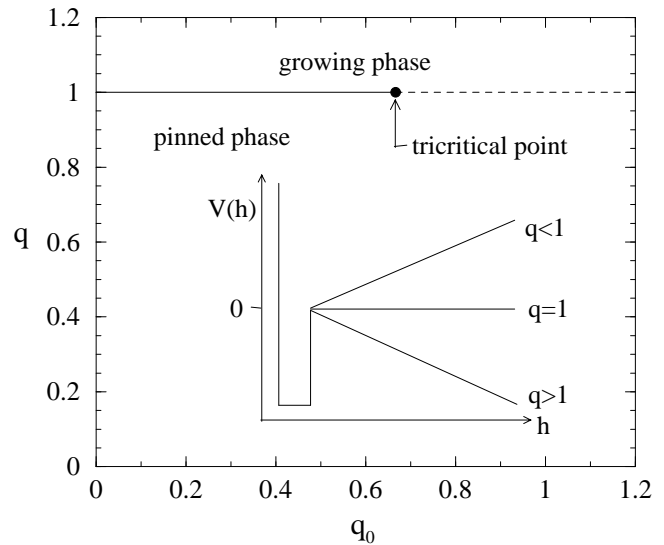}}
\caption{
\label{Fig1}
Phase diagram for $p=1$. The discontinuous (continuous)
part of the transition line is represented by a solid (dashed)
line. The inset illustrates the
potential $V(h)$ in Eq.~(5). 
}
\end{figure}
For $p=1$ and $q \leq 1$ the dynamic rules satisfy detailed
balance and the probability of finding the interface in a
configuration $\{h_1,\ldots,h_N\}$ can be expressed in
terms of a potential $V(h)$ by
\begin{equation}
\label{EquilibriumDistribution}
P(h_1,\ldots,h_N)=Z_N^{-1} \exp\biggl[-\sum_{i=1}^N V(h_i)\biggr] \ ,
\end{equation}
where the partition sum 
$Z_N=\sum_{h_1,\ldots,h_N} e^{-\sum_iV(h_i)}$
runs over all interface configurations obeying
the RSOS constraint. The potential is given by
\begin{equation}
\label{Potential}
V(h)=\left\{
\begin{array}{ll}
\infty		& \mbox{if \ } h<0 \ , \\
-\ln(q/q_0) 	& \mbox{if \ } h=0 \ , \\
-h \; \ln(q)	& \mbox{if \ } h>0 \ .
\end{array}
\right.
\end{equation} 
As shown in the inset of Fig.~\ref{Fig1}, the
attractive interaction between substrate
and bottom layer is incorporated as a
potential well at zero height.
For $q<1$ the slope of the potential 
is positive so that the interface is always pinned
to the wall. For $q>1$, where the slope is negative, 
the interface can `tunnel' through the potential barrier and
eventually detaches from the substrate. It should
be noted that in this case, the equilibrium distribution
(\ref{EquilibriumDistribution}) is no longer valid, i.e.,
the system enters a non-stationary nonequilibrium phase.

The nature of the transition depends on the depth 
of the potential well. For $q_0<\frac{2}{3}$, the potential well 
is deep enough to bind the interface to the wall at 
the transition point $q_c=1$, giving rise to a localized 
equilibrium distribution with a {\it discontinuous} transition. 
For $q_0>\frac{2}{3}$, no localized solution exists at $q=1$ and the 
transition becomes continuous. The two transition lines
are separated by a {\it tricritical} point at 
$q_0^{*}=\frac{2}{3}, q_c=1$.

In order to prove the existence of the first-order
line, we apply a transfer matrix 
formalism~\cite{TransferMatrix}. Defining a transfer
matrix $T$ acting in spatial direction by
\begin{equation}
T_{h,l}=\left\{
\begin{array}{ll}
q/q_0 	& \mbox{if \ } |h-l|\leq 1 \mbox{ \ and \ } l=0 \ , \\
q^{l} 	& \mbox{if \ } |h-l|\leq 1 \mbox{ \ and \ } l>0 \ , \\
0	& \mbox{otherwise} \ ,
\end{array}
\right.
\end{equation}
we compute the eigenvector $\phi$
of $T$ corresponding to the largest eigenvalue 
$\mu$, which determines the steady-state 
properties of the system. For $q=1$ the
solution reads
\begin{equation}
\mu	=(z+1)/q_0, \qquad
\phi_0	=q_0, \qquad
\phi_h	=z^h, 
\end{equation}
where $h\geq1$ and
\begin{equation}
z=\frac{\sqrt{1+2q_0-3q_0^2}}{2(1-q_0)}-\frac12\,.
\end{equation}
The stationary density of exposed sites 
at the bottom layer is given by 
$n_0=\phi_0^2/\sum_{h=0}^{\infty}\phi_h^2$.
It is nonzero for $q_0<\frac{2}{3}$ and vanishes 
linearly at the tricritical point. This proves the 
existence of the first-order phase transition line
in Fig.~\ref{Fig1}.

In Ref.~\cite{PreviousPaper} the density 
$n_0$ and the interface width 
$w=\langle (h-\langle h \rangle )^2 \rangle^{1/2}$
at $q_0=q$ were found to scale as
\begin{equation}
\label{Scaling1}
n_0 \sim (q_c-q)^{x_0}, \qquad
w\sim (q_c-q)^{-\gamma} \ ,
\end{equation}
with the critical exponents $x_0=1$ and $\gamma=\frac{1}{3}$. 
Using the transfer matrix approach, we can prove that these bulk
exponents remain valid along the entire second
order phase transition line, except for the tricritical
point where $x_0=\gamma=\frac13$. 
Moreover, approaching the tricritical point 
from the left along the first order transition line, 
it can be shown that the two quantities scale as
\begin{equation}
\label{Scaling2}
n_0 \sim (q_0^{*}-q_0)^{x_0^{*}}, \qquad
w\sim (q_0^{*}-q_0)^{-\gamma^{*}} \,,
\end{equation}
where $x_0^{*}=\gamma^{*}=1$.

\vspace{2mm}
\paragraph{The case $p \neq 1$: }
%
%
\begin{figure}
\epsfxsize=85mm
\centerline{\epsffile{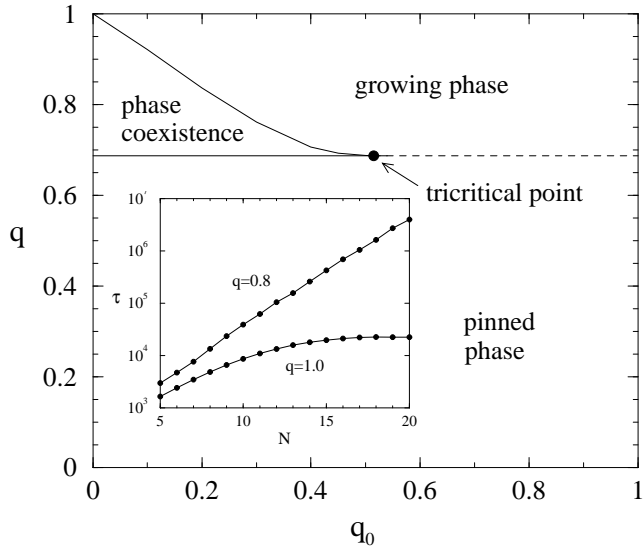}}
\caption{
\label{Fig2}
Phase diagram for $p=0.2$. The inset shows the average
time $\tau$ for the interface to detach from the wall
as a function of the system size~$N$ for $p=q_0=0.2$ 
inside and outside the coexistence regime.
}
\end{figure}
In this case the dynamic rules do not satisfy detailed balance
and the model cannot be solved using the previous methods. 
Performing Monte-Carlo simulations we determined the phase diagrams
for various values of $p$. For $p<1$ we find that the moving 
and the pinned phases {\em coexist} 
in a whole region of the parameter space rather than just on a 
line, as is the case for equilibrium first order transitions.
As shown in Fig.~\ref{Fig2},
the coexistence regime for $p=0.2$ ends at the tricritical 
point  $q_0^{*} = 0.515(10), q_c=0.6868(2)$,
where the second-order phase transition line starts.
Unlike metastable states, the pinned phase is
{\em thermodynamically stable} inside the coexistence regime, 
i.e., its life time $\tau$  grows exponentially with 
the system size, as shown in the inset of Fig.~\ref{Fig2}. 

For $p>1$, however, there is no region of phase
coexistence and the phase diagram 
is similar to that of Fig.~\ref{Fig1}. For instance,
for $p=2$ the first-order phase transition line ends
at the tricritical point 
$q_0^{*}=0.73(1)$, $q_c=1.2326(3)$. 
For $p \neq 1$, we expect that Eqs. (9)-(10) still
describe the scaling behavior at the tricritical point,
although with  different sets of critical exponents.

In order to understand the mechanism leading to phase
coexistence for $p<1$, let us consider the evolution of a large
droplet (an interval where the interface is detached
from the bottom layer) in the vicinity of the 
upper terminal point of the coexistence regime $q=1,q_0=0$. 
Because of the RSOS constraint, the growing droplet eventually
reaches an almost triangular shape with unit slope at 
the edges. The interface of the triangular 
droplet fluctuates predominantly
by diffusion of pairs of sites with equal height. Inspecting
the dynamic rules, it is easy to verify that these `landings'
of the staircase move upwards with rate~$q$ and downwards
with rate~$1$. Hence, for $q>1,q_0=0$  the droplet is 
stable with a life-time exponential in its lateral size. 
For $q>1$ and $q_0>0$, fluctuations of the bottom layer
are biased to move upwards at the edges of the droplet. 
Thus the droplet grows and the interface eventually detaches from
the bottom layer.
On the other hand, if $q_0=0$ and $q_c<q<1$, fluctuations at
the top of the triangular droplet are biased to diffuse downwards
to the edges. Therefore, the droplet shrinks at constant velocity
in a time proportional to its size,
ensuring the stability of the pinned phase.

%
%
\begin{figure}
\epsfxsize=75mm
\centerline{\epsffile{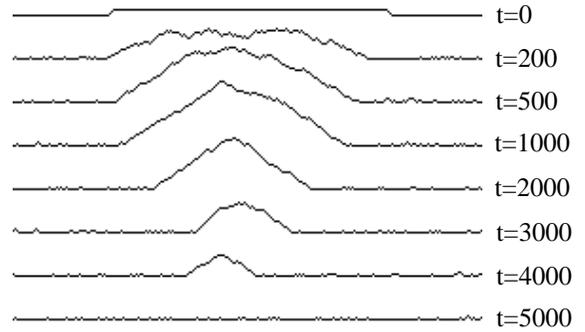}}
\vspace{4mm}
\caption{
\label{Fig3}
Temporal evolution of a partially detached interface in
the coexistence regime $q=0.8, q_0=p=0.2$. Initially
a large droplet is introduced by hand. The droplet quickly
grows, reaches a triangular shape, and finally shrinks
at constant velocity.
}
\end{figure}

As shown in Fig.~\ref{Fig3}, this robust mechanism 
for the elimination of droplets also works for positive 
values of $q_0$. If the interface detaches from the substrate over 
some distance due to fluctuations, the resulting droplet grows
and reaches an almost triangular shape. In the coexistence
regime, the droplets are biased to shrink in a time
proportional to its size, resulting in a stable pinned phase.
However, spontaneously created small islands next to the 
bottom layer contribute to the broadening of the droplets, 
reducing the range of phase coexistence. This explains why the upper
boundary of the coexistence regime decreases as $q_0$ is 
increased. At the upper boundary the
stationary density of exposed sites at the bottom layer $n_0$ 
is found to change discontinuously.

\begin{figure}
\epsfxsize=80mm
\centerline{\epsffile{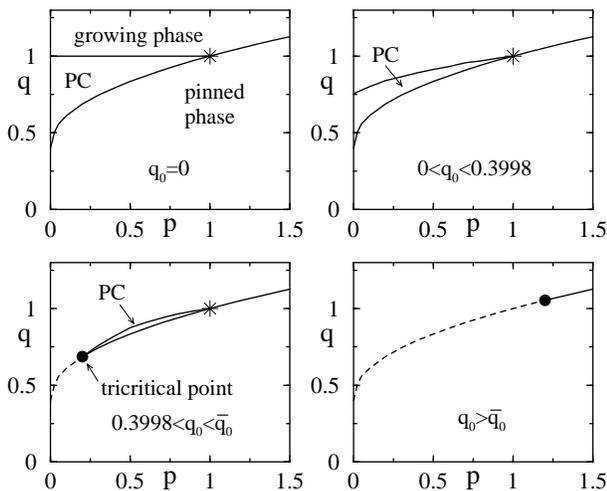}}
\caption{
\label{Fig4}
Schematic phase diagram in the $p,q$-plane,
showing the regions of phase coexistence (PC)
for fixed $q_0$.
Discontinuous and continuous transition lines
are represented by solid and dashed lines,
respectively. In the third panel the
phase coexistence regime is enlarged artificially.
The star denotes the equilibrium transition point 
(see text).
}
\end{figure}

The transition line above which the unbound phase is stable
is independent of the growth rate $q_0$. This line is the lower 
curve in Fig.~\ref{Fig4}, which is common to all four diagrams.
For $q_0$ smaller than some threshold $\bar{q}_0$, the pinned and the
unbound phases coexist in a certain region of the phase
diagram. As can be seen in the Figure, this region is bounded
by two lines which intersect to the right at the equilibrium
transition point $p=q=1$. For $q_0 < q_{c,0} \simeq 0.399$,
this is the only intersection point of the two lines and
the phase coexistence region extends down to $p=0$. On the
other hand, for $q_0 > q_{c,0}$ the two lines also intersect
on the left at another tricritical point, reducing the size of the
region of phase coexistence. This region disappears at $q_0 = \bar q_0 $.
On the basis of our numerical simulations, it is not possible
to conclude whether $\bar{q}_0 $ is equal or strictly smaller 
than $2/3$.

%
%

\vspace{2mm}
\paragraph{Discussion: }
%
Within a more general framework, the coexistence of the moving
and the pinned phase may be viewed
as follows. The evolution of the interface may be described in 
terms of the KPZ equation
\begin{equation}
\label{KPZEquation}
\partial_t h = D \nabla^2 h + 
\lambda (\nabla h)^2 +
\zeta(x,t) +
V'(h)+v_0
\end{equation}
with positive heights $h(x,t)>0$, where the velocity $v_0$ plays the
role of $q-q_c$. Clearly, for $\lambda=0$ the transition takes
place at $v_0=0$. For $\lambda>0$, the nonlinear term of 
Eq.~(\ref{KPZEquation}) may be interpreted as an additional
force acting on tilted parts of the interface in the direction
of growth. This force supports the growth of droplets wherefore
the interface detaches for any $v_0>0$. However, if $\lambda<0$
this force acts against the direction of growth. Consequently,
a sufficiently tilted interface does not propagate and may
even move downwards. For $v_0>0$ this leads 
to the formation of fluctuating droplets with a triangular shape
and a finite slope at the edges. If the short-range
force at the bottom layer is strong enough, 
such droplets, once formed, will shrink at constant velocity.
Thus, the moving and the pinned phase can only coexist in those
parts of the phase diagram where $\lambda$ is negative. In fact,
as shown in~\cite{PreviousPaper}, $\lambda$ is negative along
the transition line for $p<1$ and changes sign at $p=1$.

The phenomenon of phase coexistence was first observed
in Toom's two-dimensional north-east-center voting 
model~\cite{Toom80}. It was also
shown that open boundaries in certain one-dimensional diffusive
models may exhibit similar phenomena~\cite{Bridge}. 
The model discussed
in this work demonstrates that phase coexistence can also
emerge in {\it homogeneous} one-dimensional driven systems.

We would like to thank M.R. Evans for valuable discussions.
The support of the Israel Science Foundation, the Israel
Ministry of Science, and the Inter-University High Performance 
Computation Center is gratefully acknowledged.
H.H. would like to thank the Weizmann Institute for
hospitality where parts of this work have been done.\\[-8mm]
%
%
%
%

\end{document}